\documentclass[aps,prb,floatfix,twocolumn,superscriptaddress]{revtex4-1}

\usepackage{graphicx}
\usepackage[caption=false]{subfig}
\usepackage[english]{babel}
\usepackage{amsmath}
\usepackage{amssymb}
\usepackage{tensor}

\newcommand{\vk}{\mathbf{k}}
\newcommand{\br}{\mathbf{r}}

\newcommand{\vR}{\mathbf{R}}
\newcommand{\vA}{\mathbf{A}}
\newcommand{\vB}{\mathbf{B}}
\newcommand{\be}{\begin{eqnarray}}
\newcommand{\ee}{\end{eqnarray}}
\newcommand{\p}{\partial}

\def\ket#1{|#1\rangle}
\def\bra#1{\langle #1 |}

\begin{document}

\title{Non-Hermitian three-dimensional two-band Hopf insulator}

\author{Yan He}
\affiliation{College of Physics, Sichuan University, Chengdu, Sichuan 610064, China}
\email{heyan$_$ctp@scu.edu.cn}

\author{Chih-Chun Chien}
\affiliation{School of Natural Sciences, University of California, Merced, CA 95343, USA.}
\email{cchien5@ucmerced.edu}

\begin{abstract}
The Hopf insulator is a three-dimensional topological insulator outside the standard classification of topological insulators. Here we consider two types of non-Hermitian Hopf insulators, one without and one with the non-Hermitian skin effect. The isolated gapless points of the Hermitian model are broadened into finite regimes in the non-Hermitian models. However, the modulus of the Hopf index remains quantized in the gapped regions. The model without the non-Hermitian skin effect allows an accurate evaluation of its generalized Hopf index and energy spectrum, showing an agreement between the gapless-regime estimations from the systems with periodic- and open- boundary conditions. Near the zero-energy plane, Fermi rings can be observed whenever the Hopf index is quantized at nonzero values, and there is a bulk-boundary correspondence between the modulus of the Hopf index and the number of Fermi rings. The other model manifests the non-Hermitian skin effect in the generalized Brillouin zone and shows the skewed profiles of the bulk states. The Hopf index and energy spectrum are shown to be sensitive to the bondary condition in the presence of the non-Hermitian skin effect.
\end{abstract}

\maketitle

\section{Introduction}
The applications of topological concepts to condensed matter systems have brought us new paradigms for classifying materials~\cite{Kane_TIRev,Zhang_TIRev,Chiu2016}. While the conventional quantum mechanics deals with Hermitian systems, there have been studies on non-Hermitian systems with topological properties~\cite{Lee16,Leykam17,Wang1,ShenPRL18,Gong18,Kawabeta19}. The energy spectra and wavefunctions of non-Hermitian systems can exhibit interesting behavior due to the possible appearance of complex numbers. The bulk-boundary correspondence relating the topological invariant of the bulk and the localized edge state at the boundary has been demonstrated in Hermitian topological systems~\cite{Kane_TIRev,Zhang_TIRev,Chiu2016}. More careful analyses are usually needed for non-Hermitian systems~\cite{Lee16,KunstPRL18,Imura19,Koch19,Yokomizo19,Zhesen19} to account for the generalized Bloch band structures and non-Hermitian skin effects in order to establish the bulk-boundary correspondence. Using a many-body approach, the observation and interpretation of non-Hermitian systems may differ~\cite{Lee19}.  Moreover, the influence of non-Hermitian systems on quantum dynamics has been analyzed~\cite{SongPRL19,Yuce19}. While the Hermitian topological insulators may be classified according to their symmetries~\cite{Chiu2016}, there have been different schemes for classifying non-Hermitian topological systems~\cite{Li19,Wojcik19,Xi19,Bessho19,Zhou19,Ghatak19}.  In addition to electronic materials, non-Hermitian systems may be relevant to optical~\cite{Ganainy18}, acoustic~\cite{ZhuPRL18}, mechanical~\cite{Yoshida19,Scheibner20}, cold-atom~\cite{Gou20}, or electric-circuit~\cite{Helbig19} systems.

The Hermitian Hopf insulator is a two-band topological insulator (TI) in 3D, which is simpler than the four-band model of the AII-class TI. The ten-fold way classification of Hermitian gapped topological models is based on the stable homotopy groups~\cite{Chiu2016}. In contrast, the existence of the Hopf insulator is due to the low dimensional homotopy of the Hopf mapping~\cite{Nakahara_book} from the three-sphere $S^3$ to the two-sphere $S^2$. Therefore, the Hermitian Hopf insulator does not fit into the periodic table of topological insulators. The Hopf insulator with a unit Hopf index was introduced in Ref.~\cite{Wen}. Later, it was generalized to models with an arbitrary Hopf index~\cite{Duan}. It has been found that the number of edge states of the Hopf insulator with a higher Hopf index is usually larger than the corresponding Hopf index~\cite{Duan}, complicating the bulk-boundary correspondence. In this paper, we will only consider the model constructed by the mapping with a unit Hopf index. Since the target space of the Hopf mapping is a 2D sphere, the Hopf insulator must be a two-band model. It becomes topologically trivial when more bands are added into the model. However, the Hopf insulators were recently generalized to models with multiple bands in Ref.~\cite{Xu}. Moreover, periodically-driven systems may also exhibit the Hopf mapping~\cite{Unal19}. The Hermitian Hopf insulator has been analyzed by quantum simulators~\cite{Yuan17,Yi19}, and there may be other promising platforms for studying it~\cite{Schuster19}.

Here we investigate two generalizations of the Hopf insulator, one with diagonal non-Hermitian terms and the other with off-diagonal non-Hermitian terms. In the presence of the non-Hermitian terms, the energy spectrum may become complex. A generalization of the Hopf index for the non-Hermitian model is presented. In general, the Hopf index is complex, but its modulus exhibits quantization in the regimes similar to where the Hermitian Hopf insulator shows a quantized Hopf index. Therefore, we may still use the Hopf index to distinguish topologically distinct regimes. However, due to the non-Hermitian Hamiltonians, the gap-closing regions in the energy spectra are broadened while the Hopf index drops towards zero in those regions. This is in contrast to the Hermitian case, where the gap only closes at isolated points with the Hopf index being undefined at those points. Moreover, the broadening of gapless points into gapless regimes in other non-Hermitian models has been reported in Ref.~\cite{XuPRL17}.

By analyzing the energy spectrum of the Hopf model with diagonal non-Hermitian terms and open boundary condition along one spatial direction, we found that the gapless regimes agree with those estimated from the same system with periodic boundary condition. Moreover, there are Fermi rings at zero energy indicating  topological states crossing the band gap. A bulk-boundary correspondence relating the modulus of the generalized Hopf index and the number of Fermi rings was found in the non-Hermitian model. As we will show later, this is because the system does not exhibit the non-Hermitian skin effect. The other generalization with off-diagonal non-Hermitian terms can exhibit the non-Heritian skin effect, but its numerical results defy manageable calculations. Instead, we use approximations to extract information of the latter model. In the presence of the non-Hermitian skin effect, we will show that the energy spectrum and the Hopf index can be different if the boundary conditions are different. The sensitivity of the results to the boundary condition is a general feature of non-Hermitian topological systems~\cite{Wang1,Wang2}. The non-Hermitian Hopf insulators thus offers tractable models for studying non-Hermitian properties beyond the standard classification of topological systems.

The rest of the paper is organized as follows. Section~\ref{Sec:Theory} reviews the Hermitian Hopf insulator and presents two generalizations to the non-Hermitian setting with a generalized expression of the Hopf index. Section~\ref{sec:Result} shows the quantized values of the modulus of the Hopf index and the energy spectra of the non-Hermitian models. A bulk-boundary correspondence is established in the first non-Hermitian model. The non-Hermitian skin effect is shown to be absent in the first model but present in the second model. We also present some approximate results of the second model. Importantly, the Hopf index and energy spectrum are sensitive to the boundary condition in the presence of the non-Hermitian skin effect. Section~\ref{sec:Conclusion} concludes our work.

\section{Theoretical background}\label{Sec:Theory}
\subsection{Review of Hermitian Hopf insulator}
Following Ref.~\cite{Wen}, we construct a 3D two-band Hermitian model with a nontrivial Hopf index $\chi$ and zero Chern numbers $C_{x,y,z}$ on the three sub-2D tori. By defining
\be
&&u_1(\mathbf{k})=\sin k_x+i\sin k_y,\nonumber\\
&&u_2(\mathbf{k})=\sin k_z +i(\cos k_x + \cos k_y + \cos k_z +h), \nonumber
\ee
the 3D two-band Hamiltonian in a dimensionless form is given by
\be\label{Model}
H=\sum_{i=1}^3d_i\sigma^i,\quad d_i=\sum_{a,b=1}^2u^*_a \sigma^i_{ab} u_b.
\ee
Here $^*$ denotes the complex conjugate, $h$ is a constant parameter, and $\sigma_{i}$ with $i=1,2,3$ denotes the Pauli matrices. The Hamiltonian actually defines a map from $T^3$ to $S^2$. To see this, we can normalize $u_a$ by introducing
\be
z_a=\frac{u_a}{\sqrt{|u_1|^2+|u_2|^2}},\quad(a=1,2).
\ee
This leads to $|z_1|^2+|z_2|^2=1$, which describes a unit 3D sphere in $\mathbb{R}^4$. Thus, $z_a(\vk)$ gives a map from $T^3$ to $S^3$. We can also define the normalized $d_i$ by $R_i=\sum_{a,b=1}^2z^*_a \sigma^i_{ab} z_b$, or more explicitly by
\be
R_1=\mbox{Re}(2z_1^*z_2),\,\, R_2=\mbox{Im}(2z_1^*z_2),\quad
R_3=|z_1|^2-|z_2|^2. \nonumber
\ee
It follows that $\sum_{i=1}^3R_i^2=1$, which describes a unit 2D sphere in $\mathbb{R}^3$. Thus, $R_i$ gives a map from $S^3$ to $S^2$. This is the Hopf map~\cite{Whitehead} originally introduced by H. Hopf~\cite{Hopf31}. The composition of the above two maps gives the desired map from $T^3$ to $S^2$ with a nonzero Hopf index.

We mention that in Ref.~\cite{Duan}, the model of Eq.~(\ref{Model}) has been generalized to one with $d_i=(u^*)^p\sigma^i u^q$. Here $p$ and $q$ are integers. The generalized model gives rise to a higher Hopf index. In those generalized models, it is found that the number of edge states is usually larger than the Hopf index. For instance, Ref.~\cite{Duan} shows four surface states when the corresponding Hopf index is two. Nevertheless, we will restrict our discussion to the simplest case of Eq.~(\ref{Model}) with an established bulk-boundary correspondence and generalize it to a non-Hermitian model.

In order to give an explicit formula for the Hopf index, we first notice that the Berry curvature for a 3D two-band model can be written as
\be
F_{\mu\nu}=\frac12\vR\cdot(\p_{\mu}\vR\times\p_{\nu}\vR). \label{F1}
\ee
Here $\mu,\nu=k_x,k_y,k_z$ and $\p_{\mu}=\frac{\p}{\p k_{\mu}}$.
In terms of the variable $z_a$, the Berry curvature can be written as the curl of a globally defined vector potential~\cite{Fradkin}, given by
\be
A_{\mu}&=&\frac{i}2\sum_a\Big[z_a^*(\p_{\mu}z_a)-(\p_{\mu}z_a^*)z_a\Big],\label{A-mu}\\
F_{\mu\nu}&=&\p_\mu A_\nu-\p_\nu A_\mu \nonumber \\
&=&-i\sum_a\Big[(\p_{\mu}z_a^*)(\p_{\nu}z_a)-(\p_{\nu}z_a^*)(\p_{\mu}z_a)\Big].
\ee
The Hopf index then follows the expression
\be
\chi=\frac{1}{8\pi^2}\int_{BZ}\epsilon_{\mu\nu\rho}A_{\mu}F_{\nu\rho} d^3k.
\label{eq:HI}
\ee
Here $\epsilon_{\mu\nu\rho}$ is the Levi-Civita symbol.
For the specific model studied here, we find
\be
\chi&=&\frac{1}{2\pi^2}\int_{BZ}\frac{(s_2+h s_3)}{(3+h^2+2s_2+2hs_1)^2}d^3k; \\
s_1&=&\cos k_x+\cos k_y+\cos k_z,\nonumber\\
s_2&=&\cos k_x\cos k_y+\cos k_y\cos k_z+\cos k_z\cos k_x,\nonumber\\
s_3&=&\cos k_x\cos k_y\cos k_z. \nonumber
\ee
The Hopf index then has the following values:
\be
\chi=\left\{
       \begin{array}{ll}
         -2, & |h|<1; \\
         1, & 1<|h|<3; \\
         0, & |h|>3.
       \end{array}
     \right.
\ee
At the transition points $|h|=1$ and $|h|=3$, the dispersion becomes gapless and the two bands actually merge into one. The Hopf index is not well-defined at those transition points.

The Hopf index defined above can only characterize the homotopy of the mappings from $S^3$ to $S^2$. The actual parameter space $T^3$ contains non-contractible cycles, which make the topology of the mappings from $T^3\to S^2$ more complicated. As pointed out in Ref.~\cite{Wen}, if the system on a 2D sub-torus of $T^3$ has a non-zero Chern number, the homotopy of the mappings from $T^3\to S^2$ is a finite group rather than $\mathbb{Z}$. This case is considered in detail in Ref.~\cite{Kennedy}.
Here we verify that the complicated situation does not occur in the model shown in Eq.~(\ref{Model}). With a fixed value of $k_x$, it can be shown that the Berry curvature satisfies $F_x(k_y,k_z)=-F_x(-k_y,-k_z)$. Thus, the Chern number in this direction is zero~\cite{Duan}: $C_x=\int dk_ydk_z F_x(k_y,k_z)=0$. Similarly, we also have $C_y=0=C_z$. Therefore, the Hopf index of the model (\ref{Model}) takes values in $\mathbb{Z}$ instead of a finite group.

\begin{figure}
\includegraphics[width=\columnwidth]{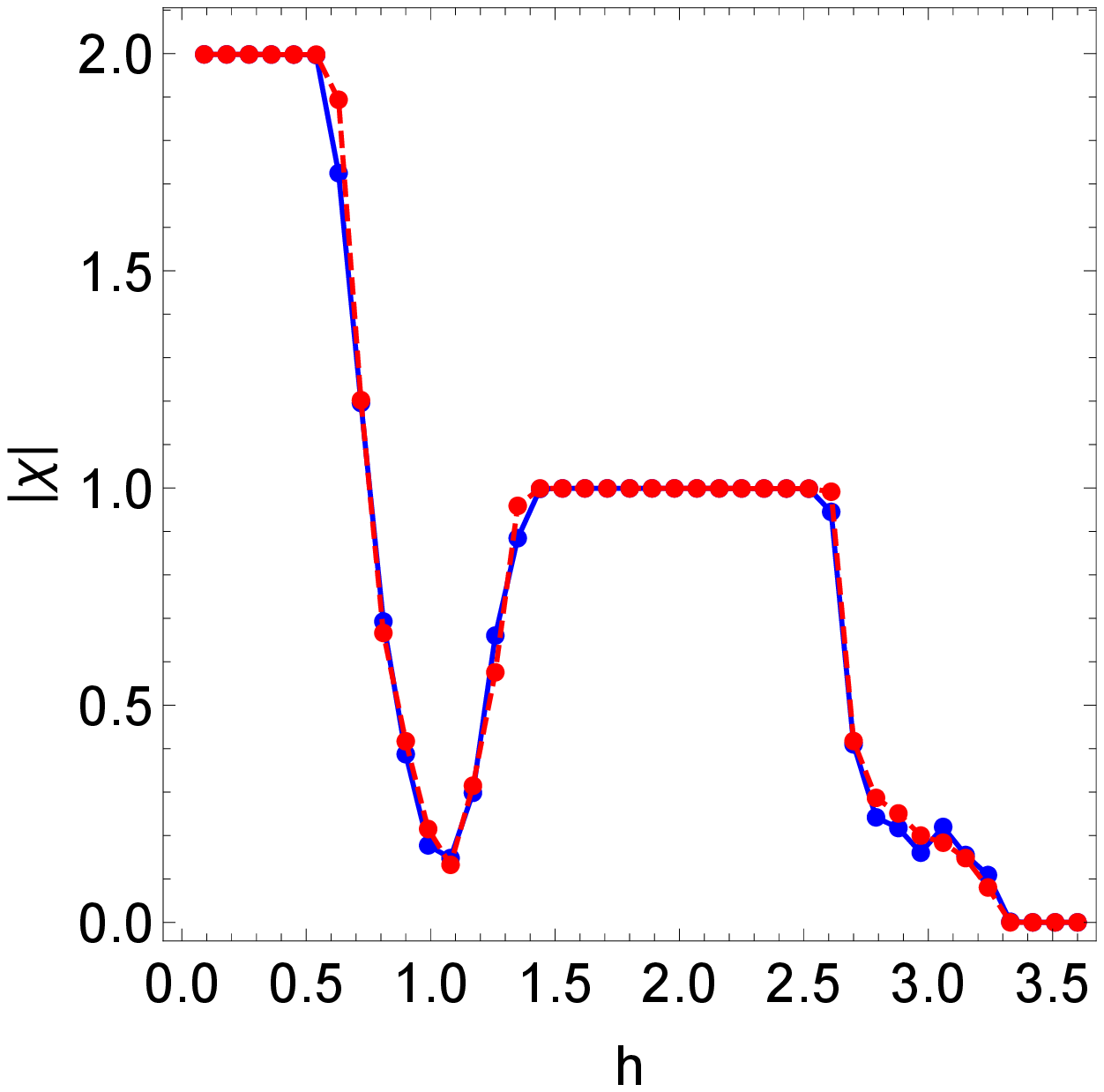}
\includegraphics[width=\columnwidth]{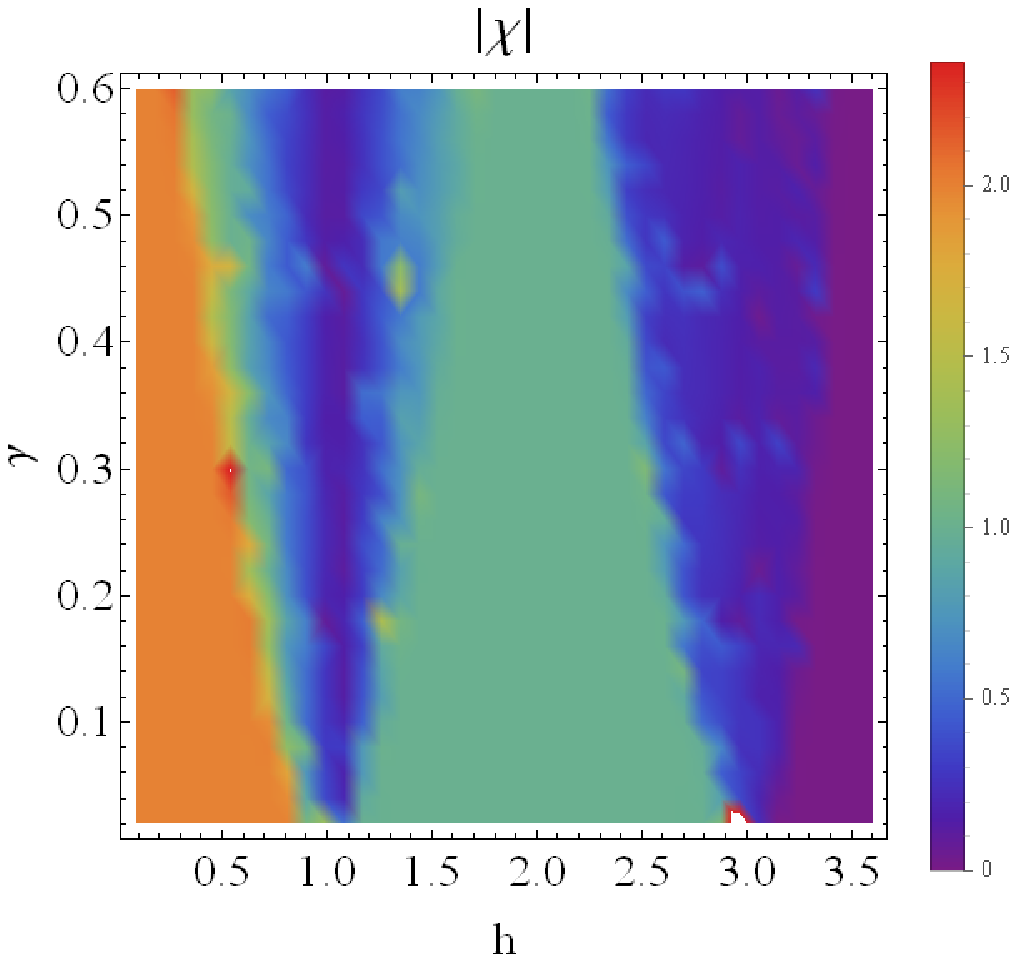}
\caption{(Top) The modulus of the Hopf index, $|\chi|$, as a function of $h$ of the model~\eqref{NH-ham} with $\gamma=0.2$. The blue solid line and the red dashed line correspond to the grid size $N=64$ and $N=128$, respectively. (Bottom) $|\chi|$ as a function of $h$ and $\gamma$. The grid size along one direction is $N=64$.}
\label{fig-H}
\end{figure}

\subsection{Non-Hermitian Hopf insulators}
Now we generalize the Hopf insulator to include non-Hermitian terms. Two generalizations will be considered here. The first one has the Hamiltonian
\be
&&H=d_1\sigma^1+d_2\sigma_2+(d_3+i\gamma)\sigma_3;\label{NH-ham}\\
&&d_1=\mbox{Re}(2u_1u_2^*),\, d_2=\mbox{Im}(2u_1u_2^*),\,
d_3=|u_1|^2-|u_2|^2. \nonumber
\ee
Here $h$ and $\gamma$ are real-valued parameters. The two eigenvalues of this Hamiltonian are
\be
E_{1,2}=\pm\sqrt{d_1^2+d_2^2+(d_3+i\gamma)^2}.
\ee
Because of the appearance of the imaginary part, the above spectrum is generally complex-valued. The model allows accurate evaluations of its energy spectrum and hopf index, but it has no non-Hermitian skin effect, as we will show shortly.

Similar to the Hermitian case, we define the following normalized vector with the components
\be
R_1=\frac{d_1}{d_0}\quad R_2=\frac{d_2}{d_0},\quad R_3=\frac{d_3+i\gamma}{d_0}.
\ee
Here $d_0=\sqrt{d_1^2+d_2^2+(d_3+i\gamma)^2}$. Note that the normalization factor $d_0$ is also a complex number in the non-Hermitian model. Although it may look natural to define the normalization factor as the norm of the vector $\mathbf{d}$, we instead choose $d_0$ to define the normalized vector $\vR$ in order to maintain the same eigenstate projectors $P_{1,2}$ as those of the the Hermitian case. Explicitly, the projectors are defined as $P_n=\ket{u_n^R}\bra{u_n^L}$ for $n=1,2$ with the left and right eigenstates satisfying $H\ket{u^R_n}=E_n\ket{u^R_n}$ and $H^{\dagger}\ket{u^L_n}=E^*_n\ket{u^L_n}$. For the two-band model, they can be expressed in terms of $\vR$ as
\be
P_{1,2}=\frac12\Big(1\pm\sum_i R_i\sigma_i\Big).\label{proj}
\ee
The Berry curvature can be expressed in terms of the projectors as
\be
F_{\mu\nu}=\p_{\mu}A_{\nu}-\p_{\nu}A_{\mu}
=i\,\mbox{Tr}\Big(P_1[\p_{\mu}P_1,\p_{\nu}P_1]\Big).
\ee
Here the Berry connection is defined as $A_{\mu}=-i\bra{u^L_n}\p_{\mu}\ket{u^R_n}$ with the left and right eigenstates.
By Eq.~(\ref{proj}), $F_{\mu\nu}$ of the non-Hermitian model is still given by the expression shown in Eq.~(\ref{F1}). We remark that in the non-Hermitian model, the components of $\vR$ are complex numbers. Thus, the Berry curvature is also complex.
It is convenient to express the Berry curvature as a 3-component dual vector
\be
B_{\rho}(\vk)=\frac{1}{2}\epsilon_{\rho\mu\nu}F_{\mu\nu}.
\ee
Here the indices $\rho,\mu,\nu$ take values of $k_x$, $k_y$, and $k_z$.

For the non-Hermitian case, it is challenging to find an explicit, analytical expression of the Berry connection $A_{\mu}$ similar to the one shown in Eq.~(\ref{A-mu}) for the Hermitian case. Nevertheless, we compute the Hopf index numerically. In order to solve the curl equation $\nabla\times \vA=\vB$, we take a Fourier transform of the Berry curvature as follows.
\be
B_{\mu}(\br)=\frac{1}{N^{3/2}}\sum_{\vk}B_{\mu}(\vk)e^{i\vk\cdot\br}.
\ee
Here $k_{x,y,z}$ take values from $\{-\pi,-\pi+\frac{2\pi}{N},\cdots,\pi-\frac{2\pi}{N}\}$, and $N$ is the number of lattice sites along one direction. Similarly, $r_{x,y,z}$ takes values from $\{-\frac{N}{2},-\frac{N}{2}+1,\cdots,\frac{N}{2}-1\}$.
The curl equation then becomes $(-i\br)\times\vA=\vB$. Under the gauge choice $\nabla\cdot\vA=0$, the Berry connection can be found as
\be
\vA(\br)=-i\frac{\br\times\vB(\br)}{r^2}.
\ee
Although the Hopf index depends on the Berry connection, it is known that the Abelian Chern-Simons term is gauge invariant up to a surface term~\cite{Nakahara_book}.
Afterwards, the Hopf index of the non-Hermitian case is given by
\be\label{eq-pchi}
\chi=-\frac{(2\pi)^3}{N^3(4\pi^2)}\sum_{\br}\vB(-\br)\cdot \vA(\br).
\ee
In principle, the generalized Hopf index is complex-valued. We mention that the Hopf index has a geometric interpretation as a linking number. The pre-images of two different points on the target space $S^2$ are two different loops in $S^3$. Those two loops link to each other and their linking number is the Hopf index. In our non-Hermitian generalization, the target space is no longer a 2D sphere in $\mathbb{R}^3$ but some hyper-surface in a complex space. Therefore, it is difficult to treat the Hopf index as a linking number following the Hermitian case. Nevertheless, we have checked that the modulus of the non-Hermitian Hopf index is quantized in a large regime of the parameter space and will present the results in the next section. In the regime where the modulus is quantized, the Hopf index seems to play the same role as that in the Hermitian case.

To explore possible non-Hermtian skin effects in generalizations of the Hopf model, we consider another model with the following Hamiltonian
\be
H_1=d_1\sigma_1+(d_2+i\gamma)\sigma_2+d_3\sigma_3\label{ham-1}.
\ee
Here the non-Hermitian term is added to the $\sigma_2$ matrix. As we will explain later, the model exhibits the non-Hermitian skin effect, but the evaluations of the energy spectrum and Hopf index are more difficult and approximations will be used.

\begin{figure}
\centerline{\includegraphics[width=\columnwidth]{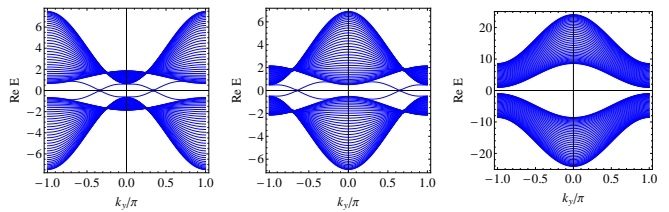}}
\caption{Energy spectrum of the non-Hermitian Hopf model~\eqref{eq-Hz} as a function of $k_y$ with open boundary along the $z$ direction. We take $k_x=2.7$ as an example. The left, middle, and right panels correspond to $h=0.2,\,1.5,\,3.8$, respectively.}
\label{fig-edge}
\end{figure}

\begin{figure}
\centerline{\includegraphics[width=\columnwidth]{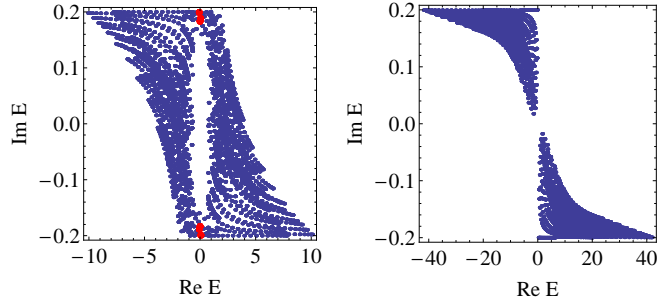}}
\caption{Energy spectrum of the non-Hermitian Hopf model~\eqref{eq-Hz} on the complex plane, showing the edge states connecting the two bands (left) and no edge state (right). The left (right) panel shows the case with $h=0.2$ and $\gamma=0.2$ ($h=3.5$ and $\gamma=0.2$). The system has open boundary condition along the $z$-direction and periodic boundary condition along the other two directions. The solid red dots on the left panel indicate the edge states.}
\label{E-complex}
\end{figure}

\begin{figure*}
\centerline{\includegraphics[width=\textwidth]{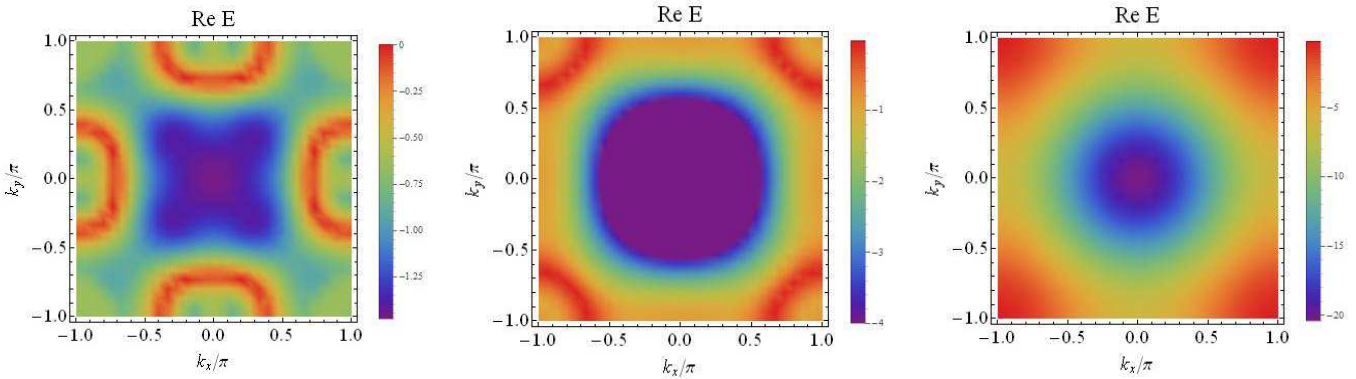}}
\caption{Energy spectrum of the eigenstate closest to the zero energy as a function of $k_x$ and $k_y$ of the model~\eqref{eq-Hz} with open boundary along the $z$ direction. From left to right, $h=0.2,\,1.9,\,3.5$, showing $2$, $1$, and $0$ zero-energy Fermi rings, respectively. We take $\gamma=0.2$.}
\label{fig-ring}
\end{figure*}

\section{Results and discussions}\label{sec:Result}
We begin with a presentation of the numerical results of the Hopf index of the non-Hermitian model~\eqref{NH-ham} with periodic boundary condition. In computing the Hopf index, we have compared the results from $N=64$ and $N=128$ grid-points. As shown in the top panel of Figure \ref{fig-H}, the two sets of data are virtually on top of each other. Thus, the grid size of $N=64$ is large enough to give stable results, and in the following we present the data with $N=64$ unless specified otherwise.

The top panel of Figure \ref{fig-H} shows the modulus of the Hopf index, $|\chi|$, as a function of $h$ for $\gamma=0.2$. One can see that $|\chi|=2$ within $0<h<0.5$ and $|\chi|=1$ within $1.5<h<2.5$. When $h>3$, the Hopf index gradually approaches zero, $|\chi|=0$. This is very similar to the case of the Hermitian Hopf insulator. However, one important difference is that the Hopf index of the non-Hermitian model is not strictly quantized around the transition points at $h=1$ and $h=3$. Instead, the Hopf index quickly drops to almost zero around those transition points. The reason is that within $0.6<h<1.4$ and $2.5<h<3.2$, the dispersion becomes gapless for the non-Hermitian model. Therefore, the Hopf index obtained numerically approaches zero. In the bottom panel of Figure \ref{fig-H}, we plot  $|\chi|$ as a function of both $h$ and $\gamma$, showing where $|\chi|=2, 1, 0$, can be found in the parameter space, respectively. The two dark blue areas around $h=1$ and $h=3$ are the gapless transition regions. As $\gamma$ increases, the gapless regions increases as well.

The ranges of the gapless regions can be determined as follows. In momentum space, the two energy bands are given by $E=\pm\sqrt{d_1^2+d_2^2+(d_3+i\gamma)^2}$. The condition for the two bands to close at $E=0$ can be expressed as
\be
d_1^2+d_2^2+d_3^2-\gamma^2=0,\quad d_3=0.
\ee
This is equivalent to the following condition
\be
|u_1|^2=\frac{\gamma}2,\quad |u_2|^2=\frac{\gamma}2. \label{u12}
\ee
For a given $\gamma$, the above two equations have real-valued solutions $\vk$ for certain ranges of the parameter $h$, which in turn determine the size of the gapless regions. It can be numerically verified that at the boundaries of the gapless regions, one always has $k_z=0$, which greatly simplifies Eq.~(\ref{u12}). The remaining two variables can be solved by requiring either $k_x=k_y$ or $k_y=\pi$. Hence, the gapless regions are given by 
\be
&&\sqrt{1-\frac{\gamma}2}-\sqrt{\frac{\gamma}2}<h<2\sqrt{1-\frac{\gamma}4}-1+\sqrt{\frac{\gamma}2}, \label{h-ga}\\
&&2+\sqrt{1-\frac{\gamma}2}-\sqrt{\frac{\gamma}2}<h<2\sqrt{1-\frac{\gamma}4}+1+\sqrt{\frac{\gamma}2}. \label{h-ga1}
\ee

After analyzing the Hopf index of the non-Hermitian Hopf insulator~\eqref{NH-ham} with periodic boundary condition, we investigate its edge states when open boundary is present. In the following, we will consider the model with open boundary condition along the $z$ direction while maintaining periodic boundary condition along the $x,y$ directions. The Hamiltonian is then given by
\be\label{eq-Hz}
&&H'=\Big[(\sin k_y-i\sin k_x)\sigma_1\nonumber\\
&&+(\sin k_x+i\sin k_y)\sigma_2-h_k\sigma_3\Big]\otimes h_0+\mbox{H.c.}\nonumber\\
&&+\Big[2h_k\sin k_y\sigma_1+2h_k\sin k_x \sigma_2\nonumber\\
&&+(\sin^2k_x+\sin^2k_y-h_k^2-1+i\gamma)\sigma_3\Big]\otimes I_0.
\ee
Here we define $h_k=h+\cos k_x+\cos k_y$, and $h_0=\delta_{i+1,j}$ and $I_0=\delta_{ij}$ are $N_z\times N_z$ matrices with $i,j=1,\cdots,N_z$.
In Figure \ref{fig-edge}, we show the energy spectrum of $H'$ as a function of $k_y$ with fixed $k_x=2.7$. From left to right, we choose $h=0.2,1.5,3.8$, respectively. One can see there are two zero-energy crossings in the left and middle panels, but there is no zero-energy crossing in the right panel. The crossings signify the edge states at the open boundary, which will be analyzed after the presentation of the energy spectrum.

Figure~\ref{E-complex} shows two typical examples of the energy eigenvalues of Eq.~(\ref{eq-Hz}) on the complex plane. Here we take $N_z=20$ points along the $z$-direction, and $26$ points along the $k_x$ and $k_y$ directions, respectively, with $\gamma=0.2$. The two examples are from $h=0.2$ and $h=3.5$, shown on the left and right panels, respectively. On the left panel, there are two separate clusters of eigenvalues corresponding to the two bands. There are also some edge states connecting those two bands, forming an enclosed, hollow region on the complex plane. On the right panel, in contrast, there are only two separate clusters of points, corresponding to the two bands. The separation of the two bands clearly shows that this model has a line gap. We found that, in general, if $|\chi|>0$ and the system is gapped, the energy spectrum is qualitatively similar to the left panel of Figure~\ref{E-complex}. In contrast, if the system is gapped with $\chi=0$, the spectrum is qualitatively similar to the right panel. However, the plots of the energy spectrum on the complex plane cannot unambiguously disclose the relation with the Hopf index.

\begin{figure}
\includegraphics[width=\columnwidth]{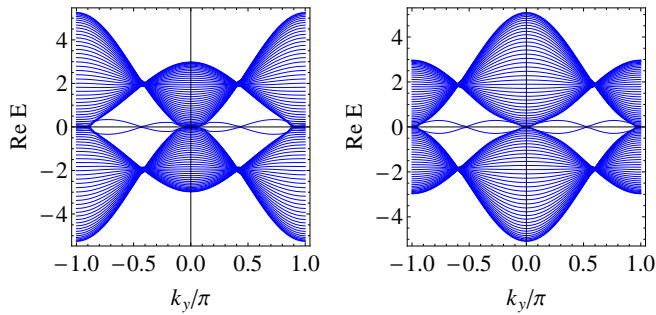}
\caption{Energy spectrum of the non-Hermitian Hopf model~\eqref{eq-Hz} as a function of $k_y$ with open boundary along the $z$ direction and $\gamma=0.2$. For the left (right) panel, $h=0.67$ and $k_x=2.84$ ($h=1.26$ and $k_x=3.04$). The gap closes in both cases.}
\label{gap-close}
\end{figure}

For a deeper understanding of the edge states crossing the energy gap, we analyze the energy spectrum as a function of $k_x$ and $k_y$. For each eigenstate, the corresponding eigen-energy forms a curved surface above the $k_x$ and $k_y$ plane. To avoid a messy view of a lot of overlapped eigen-energy surfaces, we choose to plot only the eigen-energy surface that is closest to the zero-energy and show it in Figure \ref{fig-ring}. Here we only present the part of the eigen-energy surface below the zero-energy because the part above the zero-energy has a very similar shape. We choose $h=0.2,\,1.9,\,3.5$ to represent the system with distinct values of the Hopf index. One can see that there is a ring structure with zero-energy around $\vk=(\pm\pi,\pm\pi)$ when $h=1.9$. Such a structure is known as the Fermi ring~\cite{Xu}. On the other hand, there are two Fermi rings around $\vk=(\pm\pi,0)$ and $(0,\pm\pi)$ when $h=0.2$. In contrast, there is no Fermi ring when $h=3.5$. The corresponding values of the modulus of the hopf index are $2$, $1$, and $0$, respectively. We have verified that in the gapped regimes, the number of the Fermi rings is the same as the corresponding value of the modulus of the Hopf index, thereby established a bulk-boundary correspondence for the non-Hermitian model. We remark that the bulk-boundary correspondence may be complicated or even violated in other generalizations associated with the Hopf mapping~\cite{LeeKnot18}.

\begin{figure}
    \includegraphics[width=\columnwidth]{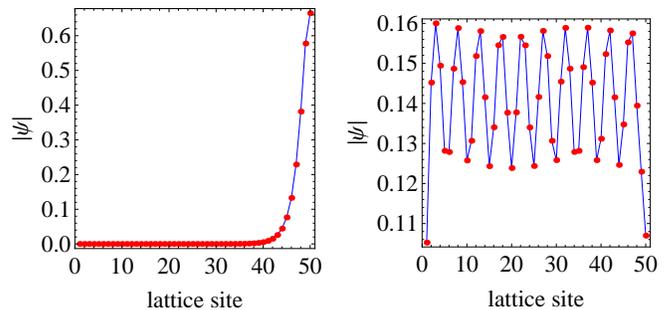}
    \caption{The amplitude $|\psi|$ of the wavefunctions of an edge state (left panel) and a selected bulk state (right panel) of the model~\eqref{eq-Hz}.  Here $h= 0.5$, $k_x= 0.2$, and $k_y= 2.3$.}
    \label{wavefunc}
\end{figure}

The qualitative feature of the Fermi ring is the same for both Hermitian and non-Hermitian Hopf insulators. The unique property of the non-Hermitian model is a finite region of $h$ that the energy spectrum is gapless for a fixed $\gamma$. In contrast, the Hermitian model is gapless only at isolated points of $h$. The size of the gapless regions have been determined by Eqs.~(\ref{h-ga}) and \eqref{h-ga1} in the calculation with periodic boundary condition. Interestingly, we can also determine the boundaries of the gapless regions of the same system with open boundary condition along the $z$ direction. In Figure \ref{gap-close}, we show the energy spectrum of the non-Hermitian model with open boundary condition along the $z$-axis and $\gamma=0.2$ as a function of $k_y$ for $h=0.67$ (left panel) and $h=1.26$ (right panel). One can clearly see that the two bulk bands close at those values of $h$. By analyzing the energy spectrum, we found that the gap remains closed when $0.67<h<1.26$. The other gapless region can be found within $2.6<h<3.3$.
One can see that the ranges of the gapless regimes determined by the open-boundary results are numerically identical to the values given by Eqs.~(\ref{h-ga}) and \eqref{h-ga1}. Thus, the gapless regimes of the non-Hermitian Hopf insulator~\eqref{NH-ham}  estimated from open- and periodic-boundary cases agree.  The agreement of the spectra of the non-Hermitian Hopf model with different boundary conditions may be because the model~\eqref{NH-ham} lacks the non-Hermitian skin effect, as we will show shortly. 
We remark that within the gapless regions, the concept of the edge states may no longer be meaningful.

To confirm the states that cross the zero-energy are localized at the open boundary in the gapped regions, Fig.~\ref{wavefunc} shows the amplitudes of the wavefunctions of an edge state and a selected bulk state of a chosen set of parameters. We have used a larger system size ($N=50$) to contrast the difference between the edge and bulk states. As shown in the left panel of Fig.~\ref{wavefunc}, the localization of the edge states is visible in the non-Hermitian Hopf model. In order to have a qualitative understanding of the edge-state wavefunction, we consider the Hermitian Hopf insulator and approximate it by a continuum model. At the Fermi ring, we have $d_3\approx 0$. Then, the Hamiltonian may be approximated by $H=d_1\sigma_1+d_2\sigma_2$. The real part of the corresponding zero-energy eigen-equation is given by
\be
(A+\frac{\p}{\p z})\psi=0
\ee
with $A=(1+h+\cos k_x+\cos k_y)$. From the approximation, we find the edge-state wavefunction to be $\psi\sim \exp(-A\,z)$, which shows an exponential decay away from the open boundary. Our numerical results suggest the edge states of the non-Hermitian model exhibits similar localization behavior.

The absence of the non-Hermitian skin effect in the model~\eqref{NH-ham} may be explained by the method discussed in Ref.~\cite{Yokomizo19}. We may treat $k_x$ and $k_y$ as two parameters and substitute $e^{ik_z}\to\beta$ into Eq.~(\ref{NH-ham}). The procedures~\cite{Yokomizo19} lead to the following generalized Bloch Hamiltonian:
\be
&&H(\beta)=d_1\sigma_1+d_2\sigma_2+(d_3+i\gamma)\sigma_3,\\
&&d_1=2(ac+bd),\, d_2=2(ad-bc),\,d_3=a^2+b^2-c^2-d^2,\nonumber\\
&&a=\sin k_x,\,b=\sin k_y,\,c=\frac{\beta-\beta^{-1}}{2i},\nonumber\\
&&d=h+\cos k_x+\cos k_y+\frac{\beta+\beta^{-1}}{2}.\nonumber
\ee
If the proper boundary equations are included, $H(\beta)$ describe an infinitely long chain with open boundary condition in the $z$ direction and periodic boundary condition along the $x$ and $y$ directions.
The eigenvalue equation becomes
\be
d_1^2+d_2^2+(d_3+i\gamma)^2=E^2.
\ee
This is a quartic equation of $\beta$, which has four complex roots $\beta_i$ for $i=1,\cdots,4$. Suppose we label those roots according to their modulus as $|\beta_1|\le|\beta_2|\le|\beta_3|\le|\beta_4|$. The continuum band can be obtained by requiring $|\beta_2|=|\beta_3|$. To obtain the eigenvalue $E$, we can solve $\beta$ from the following equation~\cite{Yokomizo19} derived by requiring $\beta_2=\beta_3e^{i\theta}$.
\be
f(\beta)=f(\beta e^{i\theta})\label{f-beta},
\ee
where $f(\beta)=d_1^2+d_2^2+(d_3+i\gamma)^2$. For a fixed $\theta$, one can solve for $\beta$ and then obtain the eigenvalues from $E^2=f(\beta)$. Due to the functional form of $f(\beta)$, Eq.~(\ref{f-beta}) can be simplified to
\be
&&c^2+d^2=(c')^2+(d')^2,\\
&&c'=\frac{\beta e^{i\theta}-\beta^{-1}e^{-i\theta}}{2i},\,
d'=\frac{\beta e^{i\theta}+\beta^{-1}e^{-i\theta}}{2}+g, \nonumber
\ee
where $g=h+\cos k_x+\cos k_y$. After some algebra, we find
\be
\beta e^{i\theta}+\beta^{-1}e^{-i\theta}=\beta+\beta^{-1},
\ee
which then implies $|\beta|=1$, regardless of the parameters. We have numerically checked that the roots of Eq.~(\ref{f-beta}) indeed satisfy $|\beta|=1$. Therefore, the non-Hermitian Hopf model shown in Eq.~(\ref{NH-ham}) has no anomalous skin effect. This because the imaginary term, $i\gamma$, only appears in front of the $\sigma_3$ matrix, which is diagonal.

\begin{figure}
\includegraphics[width=\columnwidth]{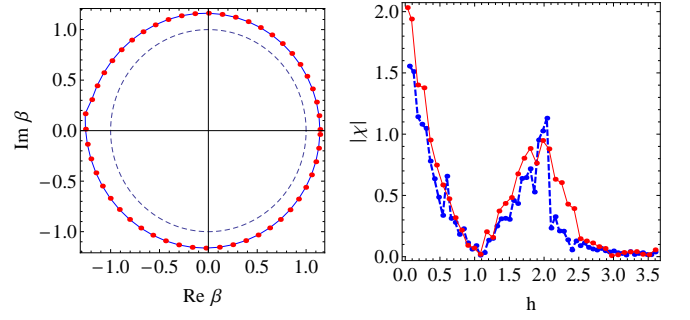}
\caption{Left panel: The generalized Brillouin zone $C_\beta$ of the model shown in Eq.~(\ref{ham-1}). Here $h=0.5$, $\gamma=0.9$, $k_x= 0.2$, and $k_y= 1.9$. The dashed curve is the unit circle for comparison. Right panel: The Hopf index of the model (\ref{ham-1}) as a function of $h$.  The thin red solid line (thick blue dashed line) shows the result from the system with open (periodic) boundary condition along the $z$ direction described by Eq.~\eqref{chi-b} (Eq.~\eqref{eq-pchi}). The system is periodic in the $x$ and $y$ directions. Here $\gamma=0.9$ and $N=64$.}
\label{beta}
\end{figure}

Next, we show the non-Hermitian skin effect of the model~\eqref{ham-1}.
To determine $\beta$ , we have to solve
\be
f_1(\beta)=f_1(\beta e^{i\theta})
\ee
with $f_1(\beta)=d_1^2+(d_2+i\gamma)^2+d_3^2$. The equation can be numerically solved, and we find $|\beta|>1$. On the left panel of Figure \ref{beta}, we plot the values of $\beta$ on the complex plane. The closed loop of $\beta$ is known as the generalized Brillouin zone \cite{Wang1}, denoted by $C_\beta$. Here we have used the parameters $h=0.5$, $\gamma=0.9$, $k_x= 0.2$, and $k_y= 1.9$. One can clear see that $C_{\beta}$ is a loop outside the unit circle and is not perfectly round. Therefore, the model shown in Eq.~(\ref{ham-1}) can exhibit non-Hermitian skin effect. The wavefunctions will be demonstrated shortly.

After the discussion of the non-Hermitian skin effect and generalized Brillouin zone $C_\beta$, it is possible to estimate the Hopf index of the model~(\ref{ham-1}) in the presence of non-Hermitian skin effect. For fixed $k_x$ and $k_y$, there is in principle a corresponding $C_\beta$ described by the equation $\beta=r(\theta)e^{i\theta}$. We can make the substitution $\exp(ik_z)\to\beta$ in the calculations of $A_{\mu}$ and $F_{\mu\nu}$. Then the Hopf index of the infinite chain along the z direction with periodic boundary condition in the x and y directions can be expressed as
\be
\chi=\frac{1}{8\pi^2}\int d^2k\int_{C_\beta}d\theta\epsilon_{\mu\nu\rho}A_{\mu}F_{\nu\rho}(e^{ik_z}\to\beta).
\label{chi-b}
\ee
When compared to 1D models, here we have to determine $C_\beta$ for all possible values of $k_x$ and $k_y$, which is very challenging. Nevertheless, one can see that $C_\beta$ is a closed loop in Fig.~\ref{beta}, with a shape similar to a circle. This feature allows $C_\beta$ to be approximated by discrete points around the loop. 

For the model (\ref{ham-1}), $C_\beta$ can be solved from the following quartic equation
\be
&&\frac{a_1}{\beta^2}+\frac{a_2}{\beta}+a_3\beta+a_4\beta^2=0.\\
&&a_1=h_k^2(1+e^{i\theta})e^{-2i\theta},\,a_4=-h_k^2(1+e^{i\theta}),\nonumber\\
&&a_2=2[\gamma(\sin k_y+i\sin k_x)+C]e^{-i\theta},\nonumber\\
&&a_3=2[\gamma(\sin k_y-i\sin k_x)-C].\nonumber
\ee
Here $C=h_k(1+\sin^2 k_x+\sin^2 k_y)+h_k^3$ and $h_k=h+\cos k_x+\cos k_y$. The numerical values of the Hopf index of Eq.~(\ref{chi-b}) for an infinitely long chain along the $z$ direction and periodic in the $x$ and $y$ directions as a function of $h$ are shown by the thin red solid line on the right panel of Figure \ref{beta}. As a comparison, we show the Hopf index calculated in momentum space~\eqref{eq-pchi} with periodic boundary condition along all directions by the thick blue dashed line. We choose $\gamma=0.9$ to be consistent with the analysis of the generalized Brillouin zone. One can see that the momentum Hopf index does not reach the quantized value $2$ as $h\rightarrow 0$ because the gapless region with periodic boundary condition extends towards $h=0$ with the large value of $\gamma$. In contrast, the open-boundary Hopf index~\eqref{chi-b} reaches the quantized value $2$ as $h\rightarrow 0$ because the gapped regime covers $h<0.09$. In the following, we will analyze the difference of the energy spectra when the boundary conditions are different.

\begin{figure}
    \includegraphics[width=\columnwidth]{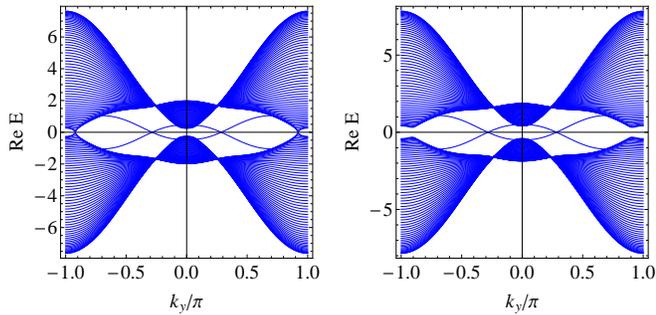}
    \caption{Energy spectrum of the model (\ref{ham-1}) with $h=0.09$ in the gapless regime (left) and $h=0.05$ in the gapped regime (right) as a function of $k_y$ with open boundary along the $z$ direction. Here $\gamma=0.9$, and $k_x=2.47$. }
    \label{ga-0.9}
\end{figure}

\begin{figure}
    \includegraphics[width=\columnwidth]{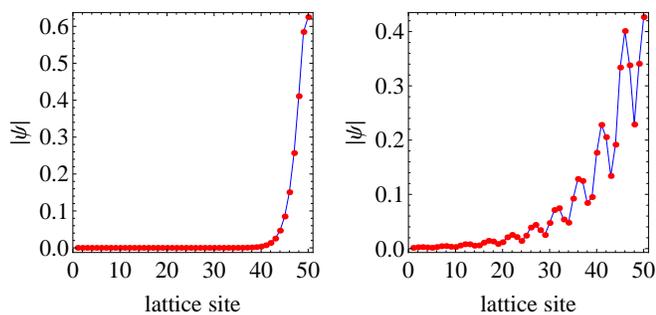}
    \caption{The modulus of the wavefunctions,  $|\psi|$, of an edge state (left panel) and a selected bulk state (right panel) of the model shown in Eq.~(\ref{eq-Hz1}) with open boundary condition.  Here $N=50$, $\gamma=0.2$, $h= 0.5$, $k_x= 0.2$, and $k_y= 1.9$.}
    \label{psi-1}
\end{figure}

The real-space Hamiltonian with open boundary condition and finite size in the $z$ direction and periodic boundary condition in the $x$ and $y$ directions is given by 
\be\label{eq-Hz1}
H_1'=H'(i\gamma\sigma_3\otimes I_0\to
i\gamma\sigma_2\otimes I_0).
\ee
Here $H'$ is from Eq.~(\ref{eq-Hz}).
Figure \ref{ga-0.9} shows the energy spectra of the non-Hermitian Hopf model~\eqref{eq-Hz}. One can see that the band gap closes at $h=0.09$ but not at smaller values of $h$. We estimate that $h=0.09$ corresponds to the boundary of the gapless region in this case. Moreover, the onset of the gapless region at $h=0.09$ when $\gamma=0.9$ agrees with the deviation of the Hopf index~\eqref{chi-b} from the quantized value if $h>0.09$, as shown in the right panel Fig.~\ref{beta}. In contrast, the energy spectrum of Eq.~\eqref{ham-1} with $\gamma=0.9$ remains gapless as $h\rightarrow 0$, causing the Hopf index~\eqref{eq-pchi} to deviate from the quantized value even when $h\rightarrow 0$. The sensitivity of the energy spectrum and Hopf index to the boundary condition provides another demonstration of the general behavior of non-Hermitian systems~\cite{Wang1,Wang2}. We remark that if the value of $\gamma$ is small, the difference between the results with different boundary conditions may be too small to be resolvable.

To illustrate the influence of the non-Hermitian skin effect on the wavefunctions, Figure \ref{psi-1} presents the modulus of of the wavefunctions, $|\psi|$, of Eq.~(\ref{eq-Hz1}).
The left and right panels show an edge state and a selected bulk state, respectively. One can see that the bulk state is also localized to the right end due to the non-Hermitian skin effects, making it harder to differentiate the bulk and edge states. We remark that although the the model \eqref{NH-ham} allows quite accurate numerical evaluations for comparing the gapless regimes, the Hopf index, and the edge states, the model \eqref{ham-1} needs numerical approximations due to the generalized Brillouin zone and higher dimensions. 
It is therefore difficult to make quantitative connections between the Hopf index from Eq.~(\ref{chi-b}) and the edge states with open boundary condition.

\section{Conclusion}\label{sec:Conclusion}
Two non-Hermitian generalizations of the 3D two-band Hopf insulator have been analyzed, one with and one without the non-Hermitian skin effect. The isolated gapless points of the Hermitian Hopf model are broadened into finite regimes with a gapless spectrum. The Hopf index has been generalized to the non-Hermitian models, and its modulus exhibits quantized values in the gapped regimes. From the energy spectrum of the model without the non-Hermitian skin effect, we found a bulk-boundary correspondence with open boundary condition between the modulus of the Hopf index and the number of  zero-energy Fermi rings. The gapless regimes estimated from the periodic- and open- boundary cases agree with each other in the absence of the non-Hermitian skin effect.

In contrast, the energy spectrum and Hopf index become sensitive to the boundary condition in the non-Hermitian Hopf model with the non-Hermitian skin effect. The edge states in the gapped regimes are shown to localized at the open boundary, but the presence of the non-Hermitian skin effect leads to skewed profiles of the bulk states and makes it more challenging to distinguish the edge and bulk states. Future refinements of the numerical approximations for the model with the non-Hermitian skin effect may reveal the full detail of the bulk-boundary correspondence. The analysis of the Hopf insulator offers an alternative view of topological systems and advance our understanding of their non-Hermitian generalizations. Moreover, the Hopf index has been studied in semimetals~\cite{Yan17,Bi17} and their non-Hermitian generalizations~\cite{YangPRB19}, showing more applications of the Hopf mapping.

\begin{acknowledgments}
Y. H was supported by the National Natural Science Foundation of China under Grant No. 11874272.
\end{acknowledgments}

\bibliographystyle{apsrev}

\begin{thebibliography}{47}
	\expandafter\ifx\csname natexlab\endcsname\relax\def\natexlab#1{#1}\fi
	\expandafter\ifx\csname bibnamefont\endcsname\relax
	\def\bibnamefont#1{#1}\fi
	\expandafter\ifx\csname bibfnamefont\endcsname\relax
	\def\bibfnamefont#1{#1}\fi
	\expandafter\ifx\csname citenamefont\endcsname\relax
	\def\citenamefont#1{#1}\fi
	\expandafter\ifx\csname url\endcsname\relax
	\def\url#1{\texttt{#1}}\fi
	\expandafter\ifx\csname urlprefix\endcsname\relax\def\urlprefix{URL }\fi
	\providecommand{\bibinfo}[2]{#2}
	\providecommand{\eprint}[2][]{\url{#2}}
	
	\bibitem[{\citenamefont{Hasan and Kane}(2010)}]{Kane_TIRev}
	\bibinfo{author}{\bibfnamefont{M.~Z.} \bibnamefont{Hasan}} \bibnamefont{and}
	\bibinfo{author}{\bibfnamefont{C.~L.} \bibnamefont{Kane}},
	\bibinfo{journal}{Rev. Mod. Phys.} \textbf{\bibinfo{volume}{82}},
	\bibinfo{pages}{3045} (\bibinfo{year}{2010}).
	
	\bibitem[{\citenamefont{Qi and Zhang}(2011)}]{Zhang_TIRev}
	\bibinfo{author}{\bibfnamefont{X.-L.} \bibnamefont{Qi}} \bibnamefont{and}
	\bibinfo{author}{\bibfnamefont{S.-C.} \bibnamefont{Zhang}},
	\bibinfo{journal}{Rev. Mod. Phys.} \textbf{\bibinfo{volume}{83}},
	\bibinfo{pages}{1057} (\bibinfo{year}{2011}).
	
	\bibitem[{\citenamefont{Chiu et~al.}(2016)\citenamefont{Chiu, Teo, Schnyder,
			and Ryu}}]{Chiu2016}
	\bibinfo{author}{\bibfnamefont{C.-K.} \bibnamefont{Chiu}},
	\bibinfo{author}{\bibfnamefont{J.~C.~Y.} \bibnamefont{Teo}},
	\bibinfo{author}{\bibfnamefont{A.~P.} \bibnamefont{Schnyder}},
	\bibnamefont{and} \bibinfo{author}{\bibfnamefont{S.}~\bibnamefont{Ryu}},
	\bibinfo{journal}{Rev. Mod. Phys.} \textbf{\bibinfo{volume}{88}},
	\bibinfo{pages}{035005} (\bibinfo{year}{2016}).
	
	\bibitem[{\citenamefont{Lee}(2016)}]{Lee16}
	\bibinfo{author}{\bibfnamefont{T.~E.} \bibnamefont{Lee}},
	\bibinfo{journal}{Phys. Rev. Lett.} \textbf{\bibinfo{volume}{116}},
	\bibinfo{pages}{133903} (\bibinfo{year}{2016}).
	
	\bibitem[{\citenamefont{Leykam et~al.}(2017)\citenamefont{Leykam, Bliokh,
			Huang, Chong, and Nori}}]{Leykam17}
	\bibinfo{author}{\bibfnamefont{D.}~\bibnamefont{Leykam}},
	\bibinfo{author}{\bibfnamefont{K.~Y.} \bibnamefont{Bliokh}},
	\bibinfo{author}{\bibfnamefont{C.}~\bibnamefont{Huang}},
	\bibinfo{author}{\bibfnamefont{Y.~D.} \bibnamefont{Chong}}, \bibnamefont{and}
	\bibinfo{author}{\bibfnamefont{F.}~\bibnamefont{Nori}},
	\bibinfo{journal}{Phys. Rev. Lett.} \textbf{\bibinfo{volume}{118}},
	\bibinfo{pages}{040401} (\bibinfo{year}{2017}).
	
	\bibitem[{\citenamefont{Yao and Wang}(2018)}]{Wang1}
	\bibinfo{author}{\bibfnamefont{S.}~\bibnamefont{Yao}} \bibnamefont{and}
	\bibinfo{author}{\bibfnamefont{Z.}~\bibnamefont{Wang}},
	\bibinfo{journal}{Phys. Rev. Lett.} \textbf{\bibinfo{volume}{121}},
	\bibinfo{pages}{086803} (\bibinfo{year}{2018}).
	
	\bibitem[{\citenamefont{Shen et~al.}(2018)\citenamefont{Shen, Zhen, and
			Fu}}]{ShenPRL18}
	\bibinfo{author}{\bibfnamefont{H.}~\bibnamefont{Shen}},
	\bibinfo{author}{\bibfnamefont{B.}~\bibnamefont{Zhen}}, \bibnamefont{and}
	\bibinfo{author}{\bibfnamefont{L.}~\bibnamefont{Fu}}, \bibinfo{journal}{Phys.
		Rev. Lett.} \textbf{\bibinfo{volume}{120}}, \bibinfo{pages}{146402}
	(\bibinfo{year}{2018}).
	
	\bibitem[{\citenamefont{Gong et~al.}(2018)\citenamefont{Gong, Ashida, Kawabata,
			Takasan, Higashikawa, and Ueda}}]{Gong18}
	\bibinfo{author}{\bibfnamefont{Z.}~\bibnamefont{Gong}},
	\bibinfo{author}{\bibfnamefont{Y.}~\bibnamefont{Ashida}},
	\bibinfo{author}{\bibfnamefont{K.}~\bibnamefont{Kawabata}},
	\bibinfo{author}{\bibfnamefont{K.}~\bibnamefont{Takasan}},
	\bibinfo{author}{\bibfnamefont{S.}~\bibnamefont{Higashikawa}},
	\bibnamefont{and} \bibinfo{author}{\bibfnamefont{M.}~\bibnamefont{Ueda}},
	\bibinfo{journal}{Phys. Rev. X} \textbf{\bibinfo{volume}{8}},
	\bibinfo{pages}{031079} (\bibinfo{year}{2018}).
	
	\bibitem[{\citenamefont{Kawabeta et~al.}(2019)\citenamefont{Kawabeta,
			Higashikawa, Gong, Ashida, and Ueda}}]{Kawabeta19}
	\bibinfo{author}{\bibfnamefont{K.}~\bibnamefont{Kawabeta}},
	\bibinfo{author}{\bibfnamefont{S.}~\bibnamefont{Higashikawa}},
	\bibinfo{author}{\bibfnamefont{Z.}~\bibnamefont{Gong}},
	\bibinfo{author}{\bibfnamefont{Y.}~\bibnamefont{Ashida}}, \bibnamefont{and}
	\bibinfo{author}{\bibfnamefont{M.}~\bibnamefont{Ueda}},
	\bibinfo{journal}{Nat. Comm.} \textbf{\bibinfo{volume}{10}},
	\bibinfo{pages}{297} (\bibinfo{year}{2019}).
	
	\bibitem[{\citenamefont{Kunst et~al.}(2018)\citenamefont{Kunst, Edvardsson,
			Budich, and Bergholtz}}]{KunstPRL18}
	\bibinfo{author}{\bibfnamefont{F.~K.} \bibnamefont{Kunst}},
	\bibinfo{author}{\bibfnamefont{E.}~\bibnamefont{Edvardsson}},
	\bibinfo{author}{\bibfnamefont{J.~C.} \bibnamefont{Budich}},
	\bibnamefont{and} \bibinfo{author}{\bibfnamefont{E.~J.}
		\bibnamefont{Bergholtz}}, \bibinfo{journal}{Phys. Rev. Lett.}
	\textbf{\bibinfo{volume}{121}}, \bibinfo{pages}{026808}
	(\bibinfo{year}{2018}).
	
	\bibitem[{\citenamefont{Imura and Takane}(2019)}]{Imura19}
	\bibinfo{author}{\bibfnamefont{K.~I.} \bibnamefont{Imura}} \bibnamefont{and}
	\bibinfo{author}{\bibfnamefont{Y.}~\bibnamefont{Takane}}
	(\bibinfo{year}{2019}), \bibinfo{note}{arXiv: 1908.09438}.
	
	\bibitem[{\citenamefont{Koch and Budich}(2019)}]{Koch19}
	\bibinfo{author}{\bibfnamefont{R.}~\bibnamefont{Koch}} \bibnamefont{and}
	\bibinfo{author}{\bibfnamefont{J.~C.} \bibnamefont{Budich}},
	\emph{\bibinfo{title}{Bulk-boundary correspondence in non-hermitian systems:
			Stability analysis for generalized boundary conditions}}
	(\bibinfo{year}{2019}), \bibinfo{note}{arXiv: 1912.07687}.
	
	\bibitem[{\citenamefont{Yokomizo and Murakami}(2019)}]{Yokomizo19}
	\bibinfo{author}{\bibfnamefont{K.}~\bibnamefont{Yokomizo}} \bibnamefont{and}
	\bibinfo{author}{\bibfnamefont{S.}~\bibnamefont{Murakami}},
	\bibinfo{journal}{Phys. Rev. Lett.} \textbf{\bibinfo{volume}{123}},
	\bibinfo{pages}{066404} (\bibinfo{year}{2019}).
	
	\bibitem[{\citenamefont{Zhang et~al.}(2019)\citenamefont{Zhang, Yang, and
			Fang}}]{Zhesen19}
	\bibinfo{author}{\bibfnamefont{K.}~\bibnamefont{Zhang}},
	\bibinfo{author}{\bibfnamefont{Z.}~\bibnamefont{Yang}}, \bibnamefont{and}
	\bibinfo{author}{\bibfnamefont{C.}~\bibnamefont{Fang}},
	\emph{\bibinfo{title}{Correspondence between winding numbers and skin modes
			in non-hermitian systems}} (\bibinfo{year}{2019}), \bibinfo{note}{arXiv:
		1910.01131}.
	
	\bibitem[{\citenamefont{Lee et~al.}(2019)\citenamefont{Lee, Lee, and
			Yang}}]{Lee19}
	\bibinfo{author}{\bibfnamefont{E.}~\bibnamefont{Lee}},
	\bibinfo{author}{\bibfnamefont{H.}~\bibnamefont{Lee}}, \bibnamefont{and}
	\bibinfo{author}{\bibfnamefont{B.~J.} \bibnamefont{Yang}},
	\emph{\bibinfo{title}{Many-body approach to non-hermitian physics in
			fermionic systems}} (\bibinfo{year}{2019}), \bibinfo{note}{arXiv:
		1912.05825}.
	
	\bibitem[{\citenamefont{Song et~al.}(2019)\citenamefont{Song, Yao, and
			Wang}}]{SongPRL19}
	\bibinfo{author}{\bibfnamefont{F.}~\bibnamefont{Song}},
	\bibinfo{author}{\bibfnamefont{S.}~\bibnamefont{Yao}}, \bibnamefont{and}
	\bibinfo{author}{\bibfnamefont{Z.}~\bibnamefont{Wang}},
	\bibinfo{journal}{Phys. Rev. Lett.} \textbf{\bibinfo{volume}{123}},
	\bibinfo{pages}{170401} (\bibinfo{year}{2019}).
	
	\bibitem[{\citenamefont{Yuce}(2019)}]{Yuce19}
	\bibinfo{author}{\bibfnamefont{C.}~\bibnamefont{Yuce}}, \bibinfo{journal}{Phys.
		Rev. A} \textbf{\bibinfo{volume}{99}}, \bibinfo{pages}{032109}
	(\bibinfo{year}{2019}).
	
	\bibitem[{\citenamefont{Li and Mong}(2019)}]{Li19}
	\bibinfo{author}{\bibfnamefont{Z.}~\bibnamefont{Li}} \bibnamefont{and}
	\bibinfo{author}{\bibfnamefont{R.~S.~K.} \bibnamefont{Mong}},
	\emph{\bibinfo{title}{Homotopical classification of non-hermitian band
			structures}} (\bibinfo{year}{2019}), \bibinfo{note}{arXiv: 1911.02697}.
	
	\bibitem[{\citenamefont{Wojcik et~al.}(2019)\citenamefont{Wojcik, Sun, Bzdusek,
			and Fan}}]{Wojcik19}
	\bibinfo{author}{\bibfnamefont{C.~C.} \bibnamefont{Wojcik}},
	\bibinfo{author}{\bibfnamefont{X.~Q.} \bibnamefont{Sun}},
	\bibinfo{author}{\bibfnamefont{T.}~\bibnamefont{Bzdusek}}, \bibnamefont{and}
	\bibinfo{author}{\bibfnamefont{S.}~\bibnamefont{Fan}},
	\emph{\bibinfo{title}{Topological classification of non-hermitian
			hamiltonians}} (\bibinfo{year}{2019}), \bibinfo{note}{arXiv: 1911.12748}.
	
	\bibitem[{\citenamefont{X et~al.}(2019)\citenamefont{X, Zhang, Gu, and
			Chen}}]{Xi19}
	\bibinfo{author}{\bibfnamefont{W.}~\bibnamefont{X}},
	\bibinfo{author}{\bibfnamefont{Z.~H.} \bibnamefont{Zhang}},
	\bibinfo{author}{\bibfnamefont{Z.~C.} \bibnamefont{Gu}}, \bibnamefont{and}
	\bibinfo{author}{\bibfnamefont{W.~Q.} \bibnamefont{Chen}},
	\emph{\bibinfo{title}{Classification of topological phases in one dimensional
			interacting non-hermitian systems and emergent unitarity}}
	(\bibinfo{year}{2019}), \bibinfo{note}{arXiv: 1911.01590}.
	
	\bibitem[{\citenamefont{Bessho et~al.}(2019)\citenamefont{Bessho, Kawabata, and
			Sato}}]{Bessho19}
	\bibinfo{author}{\bibfnamefont{T.}~\bibnamefont{Bessho}},
	\bibinfo{author}{\bibfnamefont{K.}~\bibnamefont{Kawabata}}, \bibnamefont{and}
	\bibinfo{author}{\bibfnamefont{M.}~\bibnamefont{Sato}},
	\emph{\bibinfo{title}{Topological classification of non-hermitian gapless
			phases: Exceptional points and bulk fermi arcs}} (\bibinfo{year}{2019}),
	\bibinfo{note}{arXiv: 1911.08998}.
	
	\bibitem[{\citenamefont{Zhou and Lee}(2019)}]{Zhou19}
	\bibinfo{author}{\bibfnamefont{H.}~\bibnamefont{Zhou}} \bibnamefont{and}
	\bibinfo{author}{\bibfnamefont{J.~Y.} \bibnamefont{Lee}},
	\bibinfo{journal}{Phys. Rev. B} \textbf{\bibinfo{volume}{99}},
	\bibinfo{pages}{235112} (\bibinfo{year}{2019}).
	
	\bibitem[{\citenamefont{Ghatak and Das}(2019)}]{Ghatak19}
	\bibinfo{author}{\bibfnamefont{A.}~\bibnamefont{Ghatak}} \bibnamefont{and}
	\bibinfo{author}{\bibfnamefont{T.}~\bibnamefont{Das}}, \bibinfo{journal}{J.
		Phys.: Condens. Matter} \textbf{\bibinfo{volume}{31}},
	\bibinfo{pages}{263001} (\bibinfo{year}{2019}).
	
	\bibitem[{\citenamefont{El-Ganainy et~al.}(2018)\citenamefont{El-Ganainy,
			Makris, Khajavikhan, Musslimani, Rotter, and Christodoulides}}]{Ganainy18}
	\bibinfo{author}{\bibfnamefont{R.}~\bibnamefont{El-Ganainy}},
	\bibinfo{author}{\bibfnamefont{K.~G.} \bibnamefont{Makris}},
	\bibinfo{author}{\bibfnamefont{M.}~\bibnamefont{Khajavikhan}},
	\bibinfo{author}{\bibfnamefont{Z.~H.} \bibnamefont{Musslimani}},
	\bibinfo{author}{\bibfnamefont{S.}~\bibnamefont{Rotter}}, \bibnamefont{and}
	\bibinfo{author}{\bibfnamefont{D.~N.} \bibnamefont{Christodoulides}},
	\bibinfo{journal}{Nat. Phys.} \textbf{\bibinfo{volume}{14}},
	\bibinfo{pages}{11} (\bibinfo{year}{2018}).
	
	\bibitem[{\citenamefont{Zhu et~al.}(2018)\citenamefont{Zhu, Fang, Li, Sun, Li,
			Jing, and Chen}}]{ZhuPRL18}
	\bibinfo{author}{\bibfnamefont{W.}~\bibnamefont{Zhu}},
	\bibinfo{author}{\bibfnamefont{X.}~\bibnamefont{Fang}},
	\bibinfo{author}{\bibfnamefont{D.}~\bibnamefont{Li}},
	\bibinfo{author}{\bibfnamefont{Y.}~\bibnamefont{Sun}},
	\bibinfo{author}{\bibfnamefont{Y.}~\bibnamefont{Li}},
	\bibinfo{author}{\bibfnamefont{Y.}~\bibnamefont{Jing}}, \bibnamefont{and}
	\bibinfo{author}{\bibfnamefont{H.}~\bibnamefont{Chen}},
	\bibinfo{journal}{Phys. Rev. Lett.} \textbf{\bibinfo{volume}{121}},
	\bibinfo{pages}{124501} (\bibinfo{year}{2018}).
	
	\bibitem[{\citenamefont{Yoshida and Hatsugai}(2019)}]{Yoshida19}
	\bibinfo{author}{\bibfnamefont{T.}~\bibnamefont{Yoshida}} \bibnamefont{and}
	\bibinfo{author}{\bibfnamefont{Y.}~\bibnamefont{Hatsugai}},
	\bibinfo{journal}{Phys. Rev. B} \textbf{\bibinfo{volume}{100}},
	\bibinfo{pages}{054109} (\bibinfo{year}{2019}).
	
	\bibitem[{\citenamefont{Scheibner et~al.}(2020)\citenamefont{Scheibner, Irvine,
			and Vitelli}}]{Scheibner20}
	\bibinfo{author}{\bibfnamefont{C.}~\bibnamefont{Scheibner}},
	\bibinfo{author}{\bibfnamefont{W.~T.~M.} \bibnamefont{Irvine}},
	\bibnamefont{and} \bibinfo{author}{\bibfnamefont{V.}~\bibnamefont{Vitelli}},
	\emph{\bibinfo{title}{Non-hermitian band topology in active and dissipative
			mechanical metamaterials}} (\bibinfo{year}{2020}), \bibinfo{note}{arXiv:
		2001.04969}.
	
	\bibitem[{\citenamefont{Gou et~al.}(2020)\citenamefont{Gou, Chen, Xie, Xiao,
			Deng, Gadway, Yi, and Yan}}]{Gou20}
	\bibinfo{author}{\bibfnamefont{W.}~\bibnamefont{Gou}},
	\bibinfo{author}{\bibfnamefont{T.}~\bibnamefont{Chen}},
	\bibinfo{author}{\bibfnamefont{D.}~\bibnamefont{Xie}},
	\bibinfo{author}{\bibfnamefont{T.}~\bibnamefont{Xiao}},
	\bibinfo{author}{\bibfnamefont{T.~S.} \bibnamefont{Deng}},
	\bibinfo{author}{\bibfnamefont{B.}~\bibnamefont{Gadway}},
	\bibinfo{author}{\bibfnamefont{W.}~\bibnamefont{Yi}}, \bibnamefont{and}
	\bibinfo{author}{\bibfnamefont{B.}~\bibnamefont{Yan}},
	\emph{\bibinfo{title}{Tunable non-reciprocal quantum transport through a
			dissipative aharonov-bohm ring in ultracold atoms}} (\bibinfo{year}{2020}),
	\bibinfo{note}{arXiv: 2001.01859}.
	
	\bibitem[{\citenamefont{Helbig et~al.}(2019)\citenamefont{Helbig, Hofmann,
			Imhof, Abdelghany, Kiessling, Molenkamp, Lee, Szameit, Greiter, and
			Thomale}}]{Helbig19}
	\bibinfo{author}{\bibfnamefont{T.}~\bibnamefont{Helbig}},
	\bibinfo{author}{\bibfnamefont{T.}~\bibnamefont{Hofmann}},
	\bibinfo{author}{\bibfnamefont{S.}~\bibnamefont{Imhof}},
	\bibinfo{author}{\bibfnamefont{M.}~\bibnamefont{Abdelghany}},
	\bibinfo{author}{\bibfnamefont{T.}~\bibnamefont{Kiessling}},
	\bibinfo{author}{\bibfnamefont{L.~M.} \bibnamefont{Molenkamp}},
	\bibinfo{author}{\bibfnamefont{C.~H.} \bibnamefont{Lee}},
	\bibinfo{author}{\bibfnamefont{A.}~\bibnamefont{Szameit}},
	\bibinfo{author}{\bibfnamefont{M.}~\bibnamefont{Greiter}}, \bibnamefont{and}
	\bibinfo{author}{\bibfnamefont{R.}~\bibnamefont{Thomale}},
	\emph{\bibinfo{title}{Observation of bulk boundary correspondence break-down
			in topolectrical ciruits}} (\bibinfo{year}{2019}), \bibinfo{note}{arXiv:
		1907.11562}.
	
	\bibitem[{\citenamefont{Nakahara}(2003)}]{Nakahara_book}
	\bibinfo{author}{\bibfnamefont{M.}~\bibnamefont{Nakahara}},
	\emph{\bibinfo{title}{Geometry, Topology and Physics, second edition}}
	(\bibinfo{publisher}{Taylor and Francis Group}, \bibinfo{address}{New York},
	\bibinfo{year}{2003}).
	
	\bibitem[{\citenamefont{Moore et~al.}(2008)\citenamefont{Moore, Ran, and
			Wen}}]{Wen}
	\bibinfo{author}{\bibfnamefont{J.~E.} \bibnamefont{Moore}},
	\bibinfo{author}{\bibfnamefont{Y.}~\bibnamefont{Ran}}, \bibnamefont{and}
	\bibinfo{author}{\bibfnamefont{X.-G.} \bibnamefont{Wen}},
	\bibinfo{journal}{Phys. Rev. Lett.} \textbf{\bibinfo{volume}{101}},
	\bibinfo{pages}{186805} (\bibinfo{year}{2008}).
	
	\bibitem[{\citenamefont{Deng et~al.}(2013)\citenamefont{Deng, Wang, Shen, and
			Duan}}]{Duan}
	\bibinfo{author}{\bibfnamefont{D.~L.} \bibnamefont{Deng}},
	\bibinfo{author}{\bibfnamefont{S.~T.} \bibnamefont{Wang}},
	\bibinfo{author}{\bibfnamefont{C.}~\bibnamefont{Shen}}, \bibnamefont{and}
	\bibinfo{author}{\bibfnamefont{L.~M.} \bibnamefont{Duan}},
	\bibinfo{journal}{Phys. Rev. B} \textbf{\bibinfo{volume}{88}},
	\bibinfo{pages}{201105(R)} (\bibinfo{year}{2013}).
	
	\bibitem[{\citenamefont{Liu et~al.}(2017)\citenamefont{Liu, Vafa, and Xu}}]{Xu}
	\bibinfo{author}{\bibfnamefont{C.}~\bibnamefont{Liu}},
	\bibinfo{author}{\bibfnamefont{F.}~\bibnamefont{Vafa}}, \bibnamefont{and}
	\bibinfo{author}{\bibfnamefont{C.}~\bibnamefont{Xu}}, \bibinfo{journal}{Phys.
		Rev. B} \textbf{\bibinfo{volume}{95}}, \bibinfo{pages}{161116(R)}
	(\bibinfo{year}{2017}).
	
	\bibitem[{\citenamefont{Unal et~al.}(2019)\citenamefont{Unal, Eckardt, and
			Slager}}]{Unal19}
	\bibinfo{author}{\bibfnamefont{F.~N.} \bibnamefont{Unal}},
	\bibinfo{author}{\bibfnamefont{A.}~\bibnamefont{Eckardt}}, \bibnamefont{and}
	\bibinfo{author}{\bibfnamefont{R.~J.} \bibnamefont{Slager}},
	\bibinfo{journal}{Phys. Rev. Res.} \textbf{\bibinfo{volume}{1}},
	\bibinfo{pages}{022003(R)} (\bibinfo{year}{2019}).
	
	\bibitem[{\citenamefont{Yuan et~al.}(2017)\citenamefont{Yuan, He, Wang, Deng,
			Wang, Lian, Wang, Zhang, Zhang, Chang et~al.}}]{Yuan17}
	\bibinfo{author}{\bibfnamefont{X.~X.} \bibnamefont{Yuan}},
	\bibinfo{author}{\bibfnamefont{L.}~\bibnamefont{He}},
	\bibinfo{author}{\bibfnamefont{S.~T.} \bibnamefont{Wang}},
	\bibinfo{author}{\bibfnamefont{D.~L.} \bibnamefont{Deng}},
	\bibinfo{author}{\bibfnamefont{F.}~\bibnamefont{Wang}},
	\bibinfo{author}{\bibfnamefont{W.~Q.} \bibnamefont{Lian}},
	\bibinfo{author}{\bibfnamefont{X.}~\bibnamefont{Wang}},
	\bibinfo{author}{\bibfnamefont{C.~H.} \bibnamefont{Zhang}},
	\bibinfo{author}{\bibfnamefont{H.~L.} \bibnamefont{Zhang}},
	\bibinfo{author}{\bibfnamefont{X.~Y.} \bibnamefont{Chang}},
	\bibnamefont{et~al.}, \bibinfo{journal}{Chin. Phys. Lett.}
	\textbf{\bibinfo{volume}{34}}, \bibinfo{pages}{060302}
	(\bibinfo{year}{2017}).
	
	\bibitem[{\citenamefont{Yi et~al.}(2019)\citenamefont{Yi, Yu, Sun, Xu, Chen,
			and Pan}}]{Yi19}
	\bibinfo{author}{\bibfnamefont{C.~R.} \bibnamefont{Yi}},
	\bibinfo{author}{\bibfnamefont{J.~L.} \bibnamefont{Yu}},
	\bibinfo{author}{\bibfnamefont{W.}~\bibnamefont{Sun}},
	\bibinfo{author}{\bibfnamefont{X.~T.} \bibnamefont{Xu}},
	\bibinfo{author}{\bibfnamefont{S.}~\bibnamefont{Chen}}, \bibnamefont{and}
	\bibinfo{author}{\bibfnamefont{J.~W.} \bibnamefont{Pan}},
	\emph{\bibinfo{title}{Observation of the hopf links and hopf fibration in a
			2d topological raman lattice}} (\bibinfo{year}{2019}), \bibinfo{note}{arXiv:
		1904.11656}.
	
	\bibitem[{\citenamefont{Schuster et~al.}(2019)\citenamefont{Schuster, Flicker,
			Li, Kotochigova, Moore, Ye, and Yao}}]{Schuster19}
	\bibinfo{author}{\bibfnamefont{T.}~\bibnamefont{Schuster}},
	\bibinfo{author}{\bibfnamefont{F.}~\bibnamefont{Flicker}},
	\bibinfo{author}{\bibfnamefont{M.}~\bibnamefont{Li}},
	\bibinfo{author}{\bibfnamefont{S.}~\bibnamefont{Kotochigova}},
	\bibinfo{author}{\bibfnamefont{J.~E.} \bibnamefont{Moore}},
	\bibinfo{author}{\bibfnamefont{J.}~\bibnamefont{Ye}}, \bibnamefont{and}
	\bibinfo{author}{\bibfnamefont{N.~Y.} \bibnamefont{Yao}},
	\emph{\bibinfo{title}{Realizing hopf insulators in dipolar spin systems}}
	(\bibinfo{year}{2019}), \bibinfo{note}{arXiv:1901.08597}.
	
	\bibitem[{\citenamefont{Xu et~al.}(2017)\citenamefont{Xu, Wang, and
			Duan}}]{XuPRL17}
	\bibinfo{author}{\bibfnamefont{Y.}~\bibnamefont{Xu}},
	\bibinfo{author}{\bibfnamefont{S.~T.} \bibnamefont{Wang}}, \bibnamefont{and}
	\bibinfo{author}{\bibfnamefont{L.~M.} \bibnamefont{Duan}},
	\bibinfo{journal}{Phys. Rev. Lett.} \textbf{\bibinfo{volume}{118}},
	\bibinfo{pages}{045701} (\bibinfo{year}{2017}).
	
	\bibitem[{\citenamefont{Yao et~al.}(2018)\citenamefont{Yao, Song, and
			Wang}}]{Wang2}
	\bibinfo{author}{\bibfnamefont{S.}~\bibnamefont{Yao}},
	\bibinfo{author}{\bibfnamefont{F.}~\bibnamefont{Song}}, \bibnamefont{and}
	\bibinfo{author}{\bibfnamefont{Z.}~\bibnamefont{Wang}},
	\bibinfo{journal}{Phys. Rev. Lett.} \textbf{\bibinfo{volume}{121}},
	\bibinfo{pages}{136802} (\bibinfo{year}{2018}).
	
	\bibitem[{\citenamefont{Whitehead}(1947)}]{Whitehead}
	\bibinfo{author}{\bibfnamefont{J.~H.~C.} \bibnamefont{Whitehead}},
	\bibinfo{journal}{Proc. Natl. Acad. Sci.} \textbf{\bibinfo{volume}{33}},
	\bibinfo{pages}{117} (\bibinfo{year}{1947}).
	
	\bibitem[{\citenamefont{Hopf}(1931)}]{Hopf31}
	\bibinfo{author}{\bibfnamefont{H.}~\bibnamefont{Hopf}}, \bibinfo{journal}{Math.
		Ann.} \textbf{\bibinfo{volume}{104}}, \bibinfo{pages}{637}
	(\bibinfo{year}{1931}).
	
	\bibitem[{\citenamefont{Fradkin}(2013)}]{Fradkin}
	\bibinfo{author}{\bibfnamefont{E.}~\bibnamefont{Fradkin}},
	\emph{\bibinfo{title}{Field Theories of Condensed Matter Physics}}
	(\bibinfo{publisher}{Cambridge University Press},
	\bibinfo{address}{Cambridge, UK}, \bibinfo{year}{2013}),
	\bibinfo{edition}{2nd} ed.
	
	\bibitem[{\citenamefont{Kennedy}(2016)}]{Kennedy}
	\bibinfo{author}{\bibfnamefont{R.}~\bibnamefont{Kennedy}},
	\bibinfo{journal}{Phys. Rev. B} \textbf{\bibinfo{volume}{94}},
	\bibinfo{pages}{035137} (\bibinfo{year}{2016}).
	
	\bibitem[{\citenamefont{Lee et~al.}(2018)\citenamefont{Lee, Li, Liu, Tai,
			Thomale, and Zhang}}]{LeeKnot18}
	\bibinfo{author}{\bibfnamefont{C.~H.} \bibnamefont{Lee}},
	\bibinfo{author}{\bibfnamefont{G.}~\bibnamefont{Li}},
	\bibinfo{author}{\bibfnamefont{Y.}~\bibnamefont{Liu}},
	\bibinfo{author}{\bibfnamefont{T.}~\bibnamefont{Tai}},
	\bibinfo{author}{\bibfnamefont{R.}~\bibnamefont{Thomale}}, \bibnamefont{and}
	\bibinfo{author}{\bibfnamefont{X.}~\bibnamefont{Zhang}},
	\emph{\bibinfo{title}{Tidal surface states as fingerprints of non-hermitian
			nodal knot metal}} (\bibinfo{year}{2018}), \bibinfo{note}{arXiv: 1812.02011}.
	
	\bibitem[{\citenamefont{Yan et~al.}(2017)\citenamefont{Yan, Bi, Shen, Lu,
			Zhang, and Wang}}]{Yan17}
	\bibinfo{author}{\bibfnamefont{Z.}~\bibnamefont{Yan}},
	\bibinfo{author}{\bibfnamefont{R.}~\bibnamefont{Bi}},
	\bibinfo{author}{\bibfnamefont{H.}~\bibnamefont{Shen}},
	\bibinfo{author}{\bibfnamefont{L.}~\bibnamefont{Lu}},
	\bibinfo{author}{\bibfnamefont{S.~C.} \bibnamefont{Zhang}}, \bibnamefont{and}
	\bibinfo{author}{\bibfnamefont{Z.}~\bibnamefont{Wang}},
	\bibinfo{journal}{Phys. Rev. B} \textbf{\bibinfo{volume}{96}},
	\bibinfo{pages}{041103(R)} (\bibinfo{year}{2017}).
	
	\bibitem[{\citenamefont{Bi et~al.}(2017)\citenamefont{Bi, Yan, Lu, and
			Wang}}]{Bi17}
	\bibinfo{author}{\bibfnamefont{R.}~\bibnamefont{Bi}},
	\bibinfo{author}{\bibfnamefont{Z.}~\bibnamefont{Yan}},
	\bibinfo{author}{\bibfnamefont{L.}~\bibnamefont{Lu}}, \bibnamefont{and}
	\bibinfo{author}{\bibfnamefont{Z.}~\bibnamefont{Wang}},
	\bibinfo{journal}{Phys. Rev. B} \textbf{\bibinfo{volume}{96}},
	\bibinfo{pages}{201305(R)} (\bibinfo{year}{2017}).
	
	\bibitem[{\citenamefont{Yang and Hu}(2019)}]{YangPRB19}
	\bibinfo{author}{\bibfnamefont{Z.}~\bibnamefont{Yang}} \bibnamefont{and}
	\bibinfo{author}{\bibfnamefont{J.}~\bibnamefont{Hu}}, \bibinfo{journal}{Phys.
		Rev. B} \textbf{\bibinfo{volume}{99}}, \bibinfo{pages}{081102(R)}
	(\bibinfo{year}{2019}).
	
\end{thebibliography}

\end{document}